# Experimental evidence of total absorption by a thin absorbing layer deposited on a patterned metallic surface

Ana Díaz-Rubio[1*], Alastair P. Hibbins[2], Jorge Carbonell[1] and José Sánchez-Dehesa[1]

[1] *Wave Phenomena Group, Department of Electronics Engineering, Universitat Politècnica de València, Camino de Vera S/N. (Building 7F1), ES-46022 Valencia, Spain.*

[2] *Electromagnetic and Acoustic Materials Group, Department of Physics and Astronomy, University of Exeter, Stocker Road, Exeter, EX4 4QL, United Kingdom*

**Abstract:** This work presents an experimental demonstration of total absorption by a metal-dielectric metasurface. Following a theoretical proposal [A. Díaz-Rubio et al. Phys. Rev. B 89, 245123 (2014)], we fabricated a metasurface consisting of a low absorbing dielectric layer (made of FR4) positioned on top of a metallic surface patterned with a square array of coaxial cavities. For p-polarized waves, it is observed a low frequency peak with perfect absorption. The behavior of this peak has been experimentally characterized for different dielectric layer thicknesses, coaxial cavity lengths and angles of incidence. The experimental results are in excellent agreement with numerical simulation and support the previous theoretical findings.

Electromagnetic absorbers have attracted much interest due to the number of applications in which they are involved. The design of flat and thin materials with high absorption is still a challenge. Recently, the introduction of metamaterials has opened new ways for designing absorbing materials. Metamaterials are artificial structures whose elements are arranged periodically on a subwavelength scale. By using electrically small inclusions, compared to the operation wavelength, one can create artificial metamaterials with unusual characteristics not found in natural materials [1,2]. Working under the homogenization limit, the periodic system can be considered as an effective media whose behavior can be tailored almost without limitations. These artificial materials have given the opportunity for many interesting applications like perfect lenses, optical cloaking and perfect absorbers [3-5]. If the metamaterial approach to the design of high-performance absorbers is to yield success, then any prototype device will be required to satisfy at least one of the following criteria (i) perfect or near unity absorption, (ii) very thin or sub-wavelength size to avoid bulky devices, and (iii) broadband operation.

To these purposes, a number of options have been proposed and analyzed in the literature. Structures based on resonant patches [6,7] have been studied at several spectral regimes. Usually, the resonant characteristics of these patches are employed to optimize absorption. Also, in combination with metallic backed planes, thin layers have been proposed including slits over dielectric layers [8, 9], holes or cavities in the metallic plane [10, 11] and even metal-dielectric multilayered structures [12]. Wide incidence angles can be explored with any of these possible element configurations [13].

Also, if the sub-wavelength thickness requirement is relaxed, performance improvements in broadband operation can be achieved [14, 15].

Previously, some of the current authors proposed a strongly absorbing "metasurface" for microwave frequencies consisting of a thin layer made of a low loss dielectric material on top of a patterned metallic plate [16], and we emphasized its ability to tune the frequency of the absorption peak. The metallic plate under consideration consisted of a square array of closely spaced (non-diffracting) air-filled coaxial cavities. For p- (transverse magnetic, TM) polarized waves, strong localization of electromagnetic energy in the coaxial cavities on resonance enhanced the energy absorption in the dielectric slab on top, yielding peaks of total absorption. These absorption peaks appeared at wavelengths one order of magnitude larger than the thickness of the dielectric slab and their frequency positions can be controlled by means of the cavity length. In this work, we experimentally study this system and present a practical demonstration of the results obtained theoretically. In particular, the influence of the angle of incidence is characterized; the control of the peak frequency with the cavity length is demonstrated and the effect of the dielectric thickness is considered.

A schematic representation of the structure under study is plotted in Figure 1(a) with the main parameters that define the system. Also, a photograph of one of these metasurfaces and a dielectric sheet are shown in Figure 1(b), where we can see the final appearance of the square lattice of circular coaxial cavities. The dielectric material chosen for the experimental demonstration is FR4, which is glass-reinforced epoxy laminate. At the frequencies of interest, the dielectric permittivity of this material is $\varepsilon_d = 4.1(1 + 0.025i)$ [17]. Notice that the permittivity of this material has only a small imaginary part, so it has a poor performance as an absorber at the microwaves frequencies in isolation.

We have characterized the coaxial-cavity grating with three different dielectric thicknesses, $\ell = 1.2 \text{ mm}, 1.6 \text{ mm}, 2.3 \text{ mm}$ (we have called them Slab 1, Slab 2 and Slab 3). After choosing FR4 as the dielectric, we have designed the coaxial-cavity grating for having the total absorption peak in the frequency range between 5 GHz and 10 GHz. Three different samples of the coaxial grating have been manufactured, all of them have the same cavity cross-section, but different cavity lengths ($h$). The physical dimensions of the cavity cross section are: $r_e = 4 \text{ mm}$ and $r_i = 2 \text{ mm}$ the external and internal radius respectively. The three different cavity lengths are $h = 10 \text{ mm}, 5 \text{ mm}$ and $7 \text{ mm}$ (Sample 1, Sample 2 and Sample 3 repectively). The samples are formed by 40 unit cells in the x-direction and 40 unit cells in the y-direction, and the periodicity of the array is $d = 10 \text{ mm}$. The dielectric sheet entirely covers the metallic grating and it is fixed at the corners. The dimensions of the 3 samples and 3 dielectric slabs are reported in the Table 1, giving rise to a total number of nine structures that have each been experimentally characterized.

In the experimental setup, microwave radiation impinges with an incident angle $\theta$ on the sample from a rectangular waveguide horn antenna positioned at the focus of a collimating spherical mirror for generating approximately planar waves. The horn antenna is orientated and positioned such that the electric-vector of the radiation is in the plane of incidence (i.e. p- or TM polarized), and so that the plane is incidence is parallel to the xz-plane (keeping $\phi = 0$, see Fig.1(a)). The reflected beam is collected by a second rectangular horn, also orientated so that is will detect only p-polarized radiation, and placed at the focus of a second mirror that also forms an angle $\theta$ with the normal vector of the sample surface. The angle of incidence is shifted manually by changing the position and rotation of the transmission and the detection horns. A reference measurement is needed for obtaining the absorption spectra. This reference measurement is taken by replacing the sample with an aluminum plate that has the same surface area and thickness as that of the sample (with the dielectric layer). The reflectivity is obtained for the same frequency ($f$) values with the sample and with the reference plate. Using this method, the absorption produced by the sample can be calculated as

$$A(f) = 1 - \frac{R_{\text{sample}}(f)}{R_{\text{ref}}(f)} \tag{1}$$

where $R_{\text{sample}}(f)$ and $R_{\text{ref}}(f)$ are the reflectivity spectra for the sample and the aluminum plate respectively.

The first experiment carried out characterizes a metallic grating with $h = 10$ mm and $\ell = 1.2$ mm (Slab 1 on top of Sample 1). The results of this experiment are summarized in Figure 2. In Figure 2(a), the absorption spectrum is shown when the angle of incidence is $\theta = 45°$. The blue symbols represent the experimental results and the solid red line is a finite element method model of the infinite periodic system made with Ansys HFSS [18]. The dashed red line corresponds to the absorption produced with the dielectric layer (Slab 1) covering an unpatterned perfect electric conductor.

We can see that the modeled data agrees well with the experiment except for a small frequency shift: the measured peak appears at 5.63 GHz with an absorption of 93 % and the modeled peak is at 5.57 GHz with an absorption of 98 %. The variation in the peak positions is near 1% and is due to a possible air gap between the metal grating and the dielectric layer. This air gap appears because of the defects (raised edges) produced during the manufacturing process of the coaxial cavities edges, and because of the mechanical stress of the dielectric layers, which are not completely planar. The effect of the non-perfect contact between the dielectric and the metasurface is studied later. In Figure 2 (b), the same measurement has been done for different incidence angles from $\theta = 20°$ to $\theta = 70°$. As we predicted in our previous work, the peak position does not depend on the incidence angle. However the absorption amplitude changes, since there is a maximum value for $\theta = 60°$ and for higher or lower values absorption decreases. Also, it is easy to observe that, when the incidence angle is close to the normal incidence ($\theta \approx 0$), the absorption vanishes because the impinging wave cannot excite the resonant modes in the coaxial cavities without a phase variation across the surface.

With the purpose of confirming the origin of the frequency shift, we have carried out a numerical simulation in which we include an air gap between the dielectric layer and the metal grating (see inset in Figure 3 (b)). A parametrical analysis of the air gap size, denoted by $g$, has been performed for $\theta = 45°$ using Slab 1 on top of Sample 1, and the results are reported in Figure 3. Figure 3 (a) represents the absorption spectra for different values of $g$ and shows how the existence of an air gap shifts the absorption peaks to higher frequencies. This behavior gives support our explanation since it agrees with our experimental findings. From the analysis of these results, in Figure 3 (b) the relation between the parameter $g$ and the peak position is represented, showing how the frequency increases with the air gap size. In this figure, the horizontal red line marks the frequency of the absorption peak in the experiment. According to these numerical simulations, an air gap with an averaged value of 17 μm produces the frequency shift experimentally observed. In our experimental setup, it is not possible in practice to ensure a perfect contact between the metal and the dielectric surfaces.

Also we have verified the frequency dependence of the absorption peak with the dielectric layer thickness ($\ell$). For this purpose, we have employed the grating with $h = 7$ mm and each of the three dielectric layers. Figure 4 shows the absorption spectra taken for $\theta = 45°$, the symbols represent the measurements while the solid lines illustrate the HFSS simulations. It is observed how the changes of the layer thickness produce changes in the position of peak absorption. The measured absorption peaks appear at the frequencies 7.25 GHz, 7.09 GHz and 6.92 GHz, for $\ell = 1.2$ mm, $\ell = 1.6$ mm and $\ell = 2.3$ mm, respectively. The measured results are in agreement with the numerical predictions, but the frequency shift due to the air gap can be noticed again. Note that the experimental results show an asymmetrical behavior of the absorption peaks, the left part of the absorption peaks follows the same behavior as the simulations, but the right part drops more abruptly than the predictions. This is probably due to the finite size of the sample and scattering from its edges effects. The disagreement in the amplitude of the absorption peak is less than 4%, presumably because the air layer plays an increasing role. We can see how the thickness of the dielectric layer slightly affects to the amplitude. These results demonstrate how the grating increases the absorption properties of the planar dielectric layers and how it is possible to obtain total absorption with ultra-thin layers of a low loss dielectric material.

Finally, we have also studied the dependence with the cavity length. To do that, we have employed the dielectric layer with $\ell = 1.2$ mm (Slab 1) and the three metallic samples. The corresponding spectra are shown in Figure 5. Similar to the previous case, the experimental results are compared with the HFSS simulations. The measured peaks appear at 5.62 GHz, 7.25 GHz and 9.14 GHz, for $h = 10$ mm, $h = 7$ mm and $h = 5$ mm, respectively. Also we can see the asymmetry in the resonant peaks due to the edge effects. This experiment demonstrates the capability of tuning the peak position remaining the high absorption and the dependence on the $h$ parameter whose behavior can be modeled as [16]:

$$cos(k_h h) - sin(k_h h)\chi_I = 0 \qquad (2)$$

where $k_h = \omega/c_h$ is the wave number in the coaxial cavities and $\chi_I$ is the imaginary part of a lattice sum. The reader is addressed to Ref. [16] to get comprehensive information of this expression. In Figure 5, the vertical lines mark the solution of this equation. From these results, the conducted experiments corroborate the theoretical analysis done in our previous work with the mono-mode coaxial cavity approximation. The disagreement between the solutions of the equation (2) and the experiments increases when the sample yields a resonance at higher frequencies: this is a consequence of the field distribution being modified by higher modes in the coaxial cavities

In summary, this work has reported experimental evidence of the total absorption for p-polarized waves impinging a thin absorbing layer deposited on top of coaxial-type cavities patterned on a metallic plate. The array of cavities has a sub-wavelength periodicity. The experimental data demonstrate how the resonances produced by the metallic grating enhance over 3300% the absorption in the low-loss dielectric layer, compared to the response with an unpatterned metal backing. The behavior for different incidence angles has been also studied, demonstrating our previous findings regarding the independence of the peak position and the amplitude variations. The effects observed by changing the thickness of the dielectric layer have shown that total absorption can be obtained even with ultra-thin layers. Finally, the control of the absorption peak positions with the cavity length has been also demonstrated, thus supporting the theoretical predictions.

**ACKNOWLEDGMENTS**

This work was partially supported by the Spanish Ministerio de Economia y Competitividad (MINECO) under contract TEC2010-19751 and the grant EEBB-I-1408331. We acknowledge the technical help of Benjamin Tremain.

**REFERENCES**

*Corresponding authors: andiaru@hma.upv.es; jsdehesa@upv.es

| **Dielectric slab** | *Length* (cm) | *Width* (cm) | *Thickness* (mm) |
|---|---|---|---|
| Slab 1 | 40 | 40 | 1.2 |
| Slab 2 | 40 | 40 | 1.6 |
| Slab3 | 40 | 40 | 2.3 |
| **Coaxial Grating** | $r_e$ (mm) | $r_i$ (mm) | $h$ (mm) |
| Sample 1 | 4 | 2 | 10 |
| Sample 2 | 4 | 2 | 7 |
| Sample 3 | 4 | 2 | 5 |

**Table 1:** Physical dimensions of the samples. In all the grating the periodicity is $d = 10$ mm.

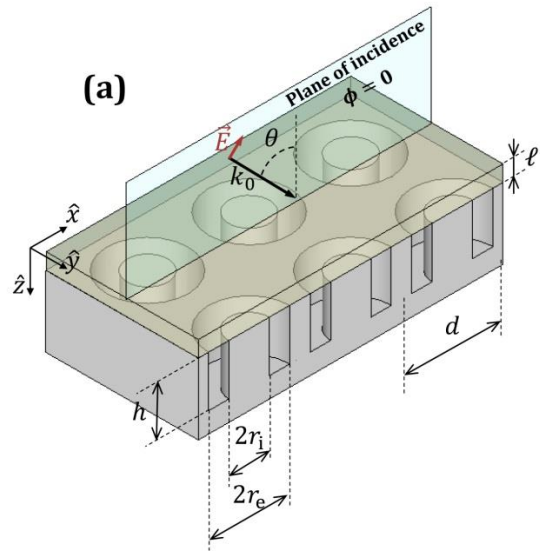

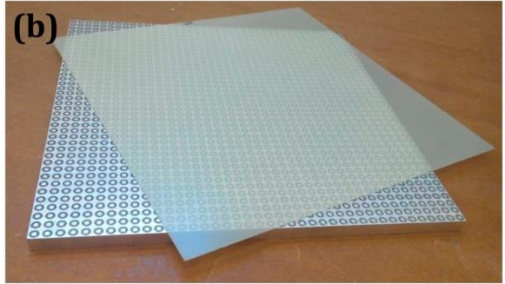

**Figure 1:** (a) Schematic representation of the experimental sample comprising of an array of coaxial cavities in a metal covered by a dielectric sheet. The plane of incidence is also shown. (b) Photograph of one constructed device. The FR4 dielectric layer is displaced for a better observation of the patterned surface ($r_i = $ **2 mm**, $r_i = $ **4 mm** and $= $ **10 mm** ).

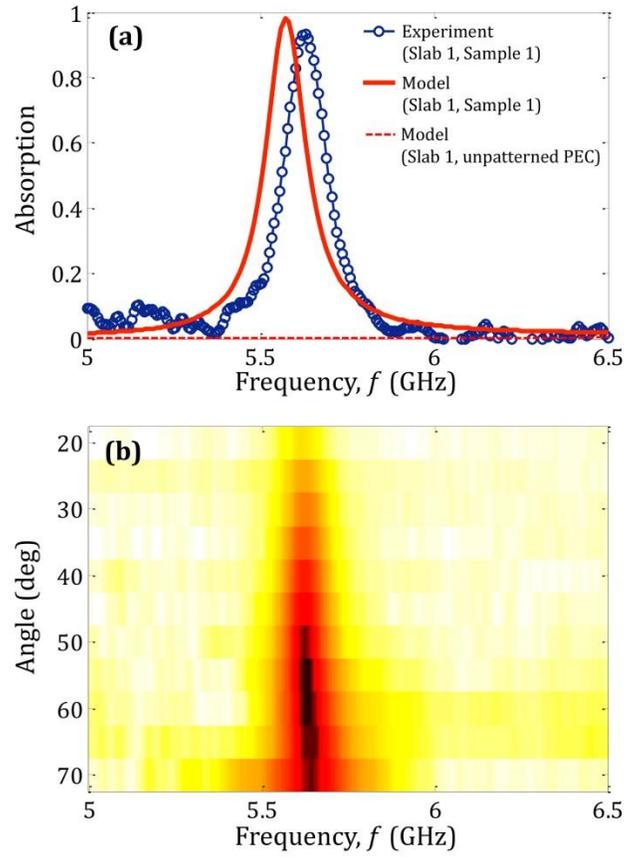

**Figure 2:** (a) Experimental and modeled (FEM) absorption spectra with $\theta = 45°$ obtained for the structure formed by Sample 1 ($h = 10$ mm) and Slab 1 ($\ell = 1.2$ mm). (b) Experimental absorption spectra as a function of the incidence angle $\theta$, where the darkest shading corresponds to the strongest absorption.

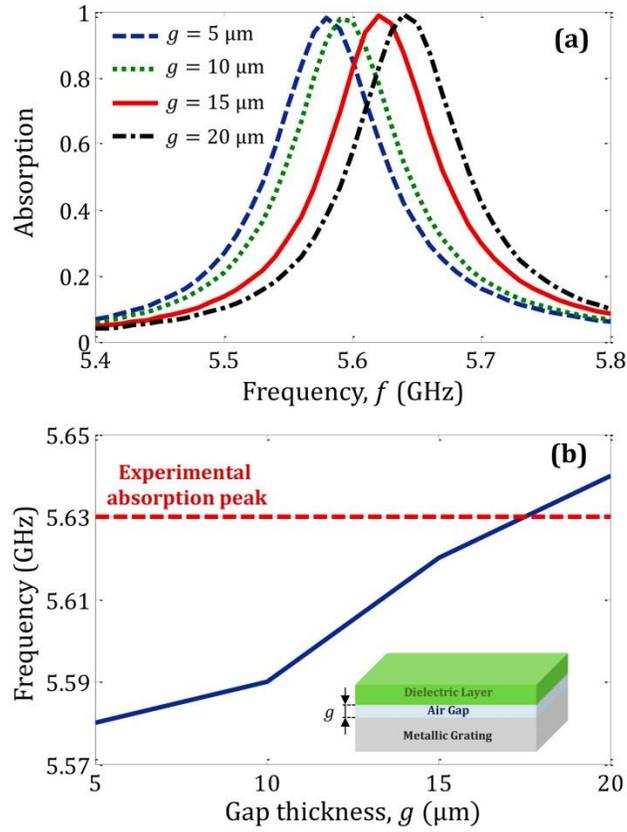

**Figure 3**: Effect of non-perfect contact between the dielectric layer and the metallic grating on the frequency of the absorption peak (numerical predictions) for the Sample 1 ($h = 10$ mm) with the Slab 1 ($\ell = 1.2$ mm) when the incidence angle is $\theta = 45°$. (a) Absorption spectra for different air gap thicknesses and (b) absorption peak position as a function of the air gap thickness, $g$. The inset illustrates the structure studied here.

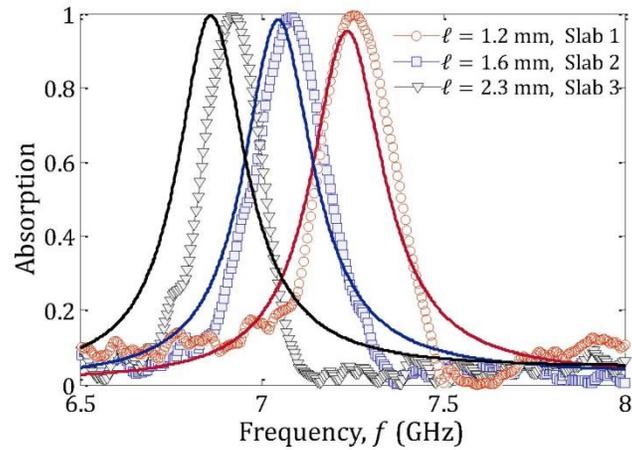

**Figure 4:** Absorption for the Sample 2 ($h = 7$ mm) with three different dielectric thicknesses. The angle of incidence is $\theta = 45°$. Symbols represent the measurements and solid lines HFSS simulations.

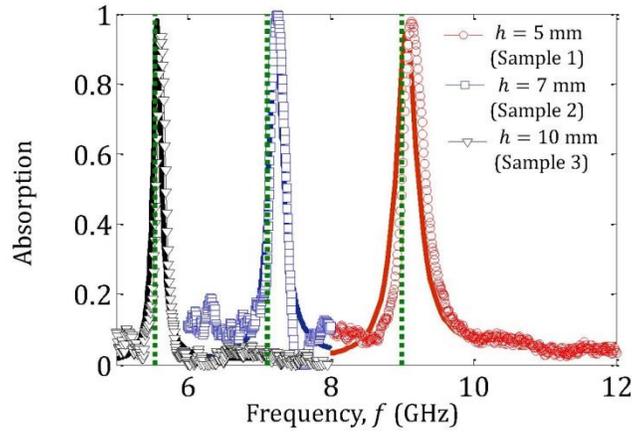

**Figure 5:** Absorption spectra for three values of cavity length $h$. The dielectric thickness is $\ell = 1.2$ **mm** (Slab 1) and the angle of incidence is $\theta = 45°$. Symbols are the measured results, solid lines represent the numerical simulations for the same value of $h$ and the vertical dashed lines represent the solutions of Equation (2).